\documentclass[prl,showpacs,amsmath,amssymb,twocolumn, 9pt]{revtex4}

\usepackage{amsthm}
\usepackage{dcolumn}
\usepackage{bm}
\usepackage{graphicx,subfigure}
\usepackage{float}

\begin{document}

\newtheorem{corollary}{Corollary}
\newtheorem{definition}{Definition}
\newtheorem{example}{Example}
\newtheorem{lemma}{Lemma}
\newtheorem{proposition}{Proposition}
\newtheorem{theorem}{Theorem}
\newtheorem{fact}{Fact}
\newtheorem{property}{Property}
\newcommand{\bra}[1]{\langle #1|}
\newcommand{\ket}[1]{|#1\rangle}
\newcommand{\braket}[3]{\langle #1|#2|#3\rangle}
\newcommand{\ip}[2]{\langle #1|#2\rangle}
\newcommand{\op}[2]{|#1\rangle \langle #2|}

\newcommand{\tr}{{\rm tr}}
\newcommand {\E } {{\mathcal{E}}}
\newcommand {\F } {{\mathcal{F}}}
\newcommand {\diag } {{\rm diag}}
\newcommand{\slocc}{\overset{\underset{\mathrm{SLOCC}}{}}{\longrightarrow}}

\title{\Large {\bf Quantum random number generator based on quantum tunneling effect }}
\author{Junlin Li$'$}
\author{Haihan Zhou$'$}
\author{Dong Pan}
\author{Guilu Long}
\email{gllong@tsinghua.edu.cn}
\affiliation{$^1$State Key Laboratory of Low-Dimensional Quantum Physics, Tsinghua University, Beijing 100084, China}
\begin{abstract}
In this paper, we proposed an experimental implementation of quantum random number generator(QRNG) with inherent randomness of quantum tunneling effect of electrons. We exploited InGaAs/InP diodes, whose valance band and conduction band shared a quasi-constant energy barrier. We applied a bias voltage on the InGaAs/InP avalanche diode, which made the diode works under Geiger mode, and triggered the tunneling events with a periodic pulse. Finally, after data collection and post-processing, our quantum random number generation rate reached 8Mb/s, and final data was verified by NIST test and Diehard test. Our experiment is characterized as an innovative low-cost, photonic source free, integratable or even chip-achievable method in quantum random number generation.        
\end{abstract}
 

\maketitle
\section{\uppercase\expandafter{\romannumeral1}. Introduction} Random numbers are crucial in many fields, for instance, the physical simulation\cite{gentle2003simulating}, information processing\citep{emerson2003pseudo},quantum communication protocols\cite{PhysRevA.68.042317}, quantum cryptography\cite{bennett1992experimental} and quantum computation\cite{nielsen2002quantum}. Under most circumstances, security is directly associated with its unpredictability and uncopyablity. However, the prevalent pseudo-random number or chaotic random number\cite{stojanovski2001chaos} are theoretically pre-determined or not proven to be unpredictable. Meanwhile, it is of great significance to develop true random number generator. Fortunately, uncertainty is a fundamental property of quantum mechanics. So, numerous studies, on true random number generation, has focused on the application of quantum inherent uncertainty or probability , such as the path choice of single photon with fixed polarization after passing a PBS\citep{rarity1994quantum}\citep{stefanov2000optical}; the uncertainty of the arrival time of single photons\cite{nie2014practical}\citep{wahl2011ultrafast}; or the phase fluctuation of photons\cite{qi2010high}\cite{xu2012ultrafast}. And more recently, Bowels proposed a protocol of self-testing quantum random number generator\cite{lunghi2015self} with the measurement of 'dimension witness'\cite{bowles2014certifying}. It made a quantative analysis on the true randomness of a given system. Also, Ma studied how to generate true randomness with an untrustworthy random source\cite{PhysRevX.6.011024}. Moreover, Xu introduced a robust quantum random number generator via the high dimensional interference\cite{xu2016experimental}. The generation speed of these protocols varies from bps to Gbps. Noteworthy, all these pervasive protocols exploited photonic sources, or, even single-photon sources.

In light of the difficulty of integration and the vulnerability to the environmental influence of photonic source, together with other flaws that impede the pragmatic application of quantum random number generators, we focused on another intrinsic randomness of quantum mechanics---the tunneling probability of electrons\citep{caldeira1981influence}\citep{banerjee2008quantum}\citep{schwartz1985quantitative}, which aborted the essence of photonic source and turn to the electronic source. Consequently, our QRNG could be highly integratable.

In this paper, we introduced an efficient protocol of quantum random number generation via the application of intrinsic indeterministic property of quantum tunneling effect and experimentally realized this protocol via the widely applied InGaAs/InP avalanche diode\cite{kanbe1980ingaas}\citep{renker2006geiger}\citep{aull2002geiger}, with the generation speed reaching 8Mb/s. Furthermore, higher speed, up to 20Mb/s, can be reached by changing the frequency of trigger pulses. On the other hand, a more efficient system which could respond to a higher frequency of trigger pulse is competent to augment the generation speed, as stable high frequency voltage pulses up to Gb/s could be realized in a precise way in nowadays electronic controlling. Also, post-processing program can be easily transplanted into our data-collecting FPGA, which enable a real-time output of quantum random number sequence.

\section{\uppercase\expandafter{\romannumeral2}. Protocol} Consider electrons trapped in a potential well, we apply a periodic bias voltage, whose peak value is $U_H$ on this system, which could induce tunneling events with a constant probability $p$ within each single period, and then record a sequence of these signal with '0' when no tunneling occurs in a single pulse period and '1' when it occurs. Here, the tunneling probability $p$ can be determined theoretically \cite{moll1964physics}. Finally, post-processing of the sequence was operated and we obtained the eligible random number sequences\cite{ma2013postprocessing}. This protocol is also summarized in Fig\ref{fig1}: 
\begin{figure}
\includegraphics[width=3in]{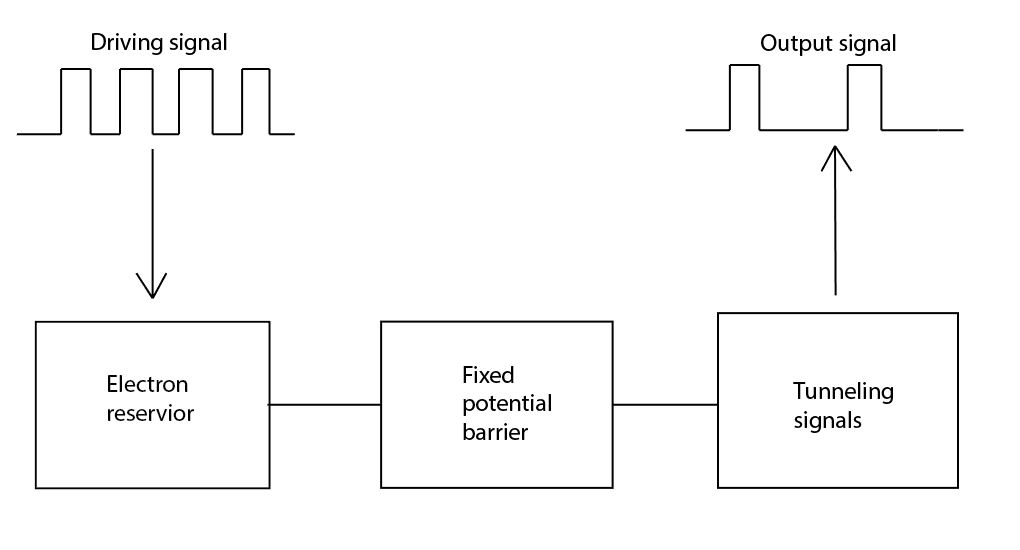}
\caption{Brief summary of the tunneling-based QRNG}
\label{fig1}
\end{figure}

We noticed that Shelan Khasro Tawfeeq has exploited the dark counts of InGaAs/InP avalanche diode in random number generation\cite{tawfeeq2009random}. However, our protocol is utterly distinct with hers. This difference is interpreted in the next section.
\section{\uppercase\expandafter{\romannumeral3}. Experimental implementation}

The key problem in our experiment is setup of the electron reservoir, bounded by a stable potential barrier, as the electronic realization of precise bias voltage pulses is not a challenge. After several pre-tests, we utilized the InGaAs/InP avalanche diode. Although the InGaAs/InP avalanche diode is prevailing in the photon detectors, our experiment is totally irrelevant to photonic source. On the contrary, we just take advantage of the quasi static barrier it possessed. As concoluded in \cite{susa1980new}, an InGaAs/InP avalanche diode consists four parts in its energy band diagram. And in our experiment,  trigger signals were applied to accelerate electrons in the $P^+-InP$ section. These electrons tunneled through junction between $P^+-InP$ section and $n-InP$ section with certain probability determined on the peak voltage of trigger signals $U_H$. Subsequently, we recorded the tunneling signals and came to raw data.

\begin{figure}
\includegraphics[width=3in]{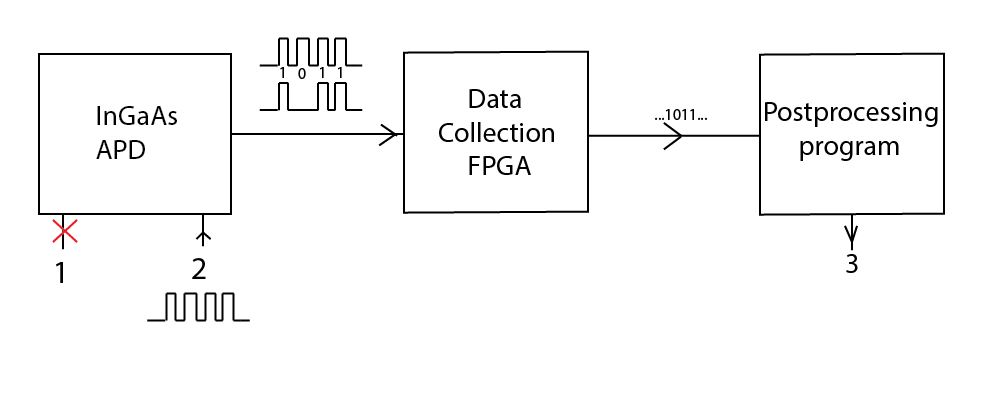}
\caption{Experimental setup of tunneling-based QRNG. $1$: Optical input channel; $2$: External clock input channel(trigger signal input channel);$3$: Final random number output channel.}
\label{fig2}
\end{figure}

\begin{figure}
\includegraphics[width=4in]{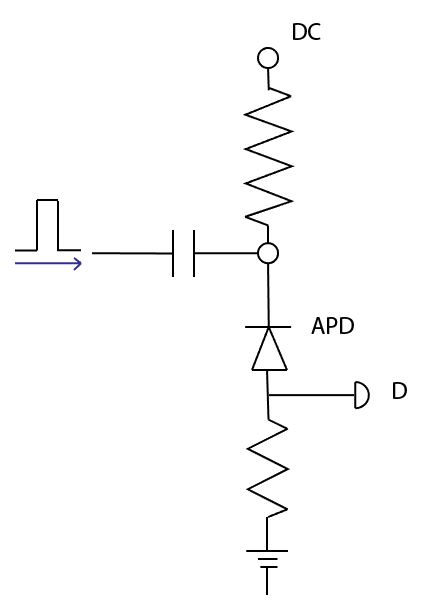}
\caption{Circuit of tunneling-based QRNG.}
\label{fig3}
\end{figure}

Setup of our experiment was shown in Fig.\ref{fig2}. And \ref{fig3} showed the circuit of out experiment. We confined the InGaAs/InP diode in a seal box. Hence, no environmental photons could contribute to the signals received by the receptor module. 

As we mentioned above, the QRNG source is the tunneling effect of electrons in InGaAs/InP avalanche diode, which is irrelevant to the photons and was considered as part of the dark counts in the previous study\cite{tosi2009ingaas}.

Dark counts of InGaAs/InP avalanche diode can be characterized into 3 kinds\cite{ribordy1998performance}

\textit{1---} Dark counts induced by heat motion of electrons.

\textit{2---} Dark counts induced by quantum tunneling effect.

\textit{3---} Dark counts induced by after-pulse effect.

When proper bias voltage is applied on the InGaAs/InP avalanche diode, it works under the Geiger mode\cite{renker2006geiger}. Under this circumstance, the accelerated electrons triggered avalanche effect in the 'accelerating section', which could induce current signals. While the bias voltage varies, the tunneling probability changes, so does the data properties(data entropy, data auto-correlation parameter and the final data generation speed).

And as mentioned in the previous section, our protocol focus more on the 'Dark counts induced by quantum tunneling effect', where the voltage of the 
trigger signal's high level $U_H$ dominates the tunneling probability in a single period $T$. While Shelan's work emphasized more about the pulsewidth of a fixed trigger signal. The after-pulse effect could responsible for her thesis, which is not credited with quantum property.

In order to restrain the $fisrt$ effect mentioned above, the working environment is monitored at $200K$ by the semiconductor cooling system. Under this circumstance, the $\uppercase\expandafter{\romannumeral1}$ type dark counts was reduced to $500/s$, which is equivalent to $10^{-5}/pulse$ in a $50Mhz$ bias voltage pulse triggered system. Meanwhile, the $\uppercase\expandafter{\romannumeral3}$ type dark counts is partly circumvented by the deadtime of this system. The deadtime system ensures that after each tunneling occurs, there will be a time interval $\Delta T$ during which the detector is forced offline. Namely, the after-pulse during this time interval $\Delta T$ could not be counted. However, due to the hardware impediment and the remnant after-pulse, our raw random number data requires a further optimization, and post-processing program is applied to countervail this bias and subsequently generate true randomness.

The counting number of the tunneling-induced signals in $1s$ can be directly displayed on a screen. In our experiment, the frequency of bias voltage pulses was set to $50Mb/s$, then we adjusted the amplitude of these pulses $U$ until the counting number reach $2.5\times 10^7$. According to \citep{RN1}, we can demonstrate our tunneling probability as :

\begin{align}
\centering
 P(V) = Ae^{\frac{-B}{V-V_0}}
\end{align} 

\begin{figure}
\includegraphics[width=3in,height = 1.7in]{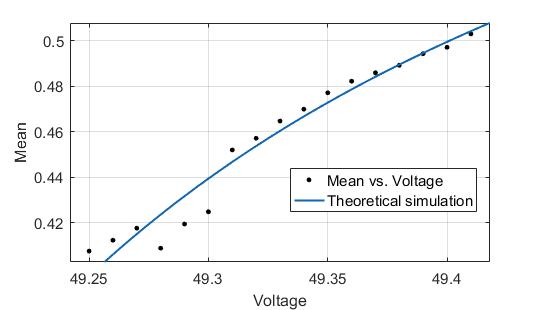}
\caption{Mean of output data under different voltage. $R^2 = 0.964$}
\label{fig4}
\end{figure}

\begin{figure}
\includegraphics[width=3in,height = 1.7in]{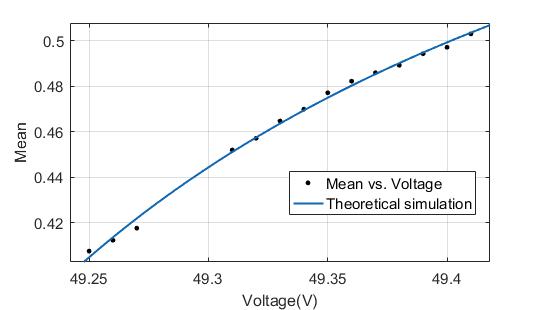}
\caption{Mean of output data under different voltage(without $49.28V, 49.29V$ and $49.30V$). $R^2 = 0.9996$}
\label{fig5}
\end{figure}

\begin{figure}
\includegraphics[width = 3in,height = 1.7in]{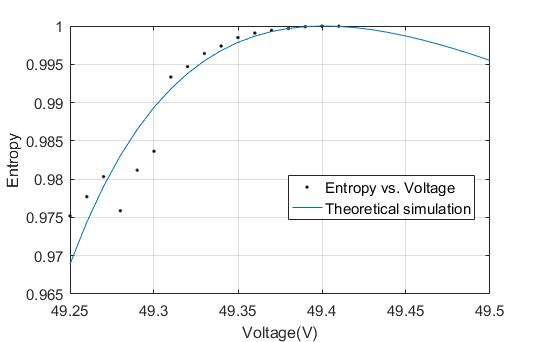}
\caption{Entropy of output data under different voltage.}
\label{fig6}
\end{figure}

\begin{figure}
\includegraphics[width=3in,height = 1.7in]{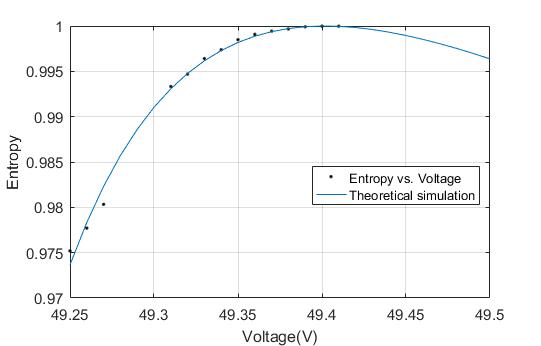}
\caption{Entropy of output data under different voltage(without $49.28V, 49.29V$ and $49.30V$).}
\label{fig7}
\end{figure}

Here, $A$,$B$ are parametric expression, which are determined by several indexes; $V_0$ is the critical voltage, under which the tunneling probability is $0$. In order to determine proper voltage, we measured the mean and entropy of output data from $49.25V$ to $49.5V$, and simulated the result based on the above equation. We obtained figure \ref{fig4},\ref{fig5},\ref{fig6},\ref{fig7}.  

We noticed when $U = 49.28V, U = 49.29V, U = 49.30V$, the mean obviously biased from other data, so we omitted them and comes to figure \ref{fig5},\ref{fig7}.
 
As the forward tunneling probability is too complicated to have an analytical expression\cite{moll1964physics}, we quantitatively fitted the data with curves shown above.

Finally, we chose $U_H = 49.40V$. And we designed a $FPGA$ module to collect all tunneling data and saved it into a $.txt$ document. The speed of raw data collection is about $20Mb/s$, which is restricted by the $USB$ communication serial port.

\section{\uppercase\expandafter{\romannumeral4}. Post-processing}
The post-processing program is realized by the application of Toeplitz-hashing extractor\cite{ma2013postprocessing}.

\textit{Min-entropy estimation---} We measure the minimum entropy $H_m$ of our raw data\cite{konig2009operational}. 
\begin{equation}
H_m = -log_2(\underset{x}{max} P[x])
\end{equation}
Here $x$ refers to all the possible sequence that ${0,1}^n$ could reach. In our scheme, we took $n = 8$, namely, we divide our raw data into $8-bits$ sequences. And then calculate maximum probability of these segments and the min-entropy. In our experiments, $H_m = 5.1204$.

\textit{Toeplitz-hashing extractor---} After the min-entropy estimation, we characterized the raw data with the proportion of quantum randomness. Namely, independent $2.8-bits$ quantum random code can be extracted from each $8-bits$ raw data segment. Subsequently, we generate a Toeplitz matrix $T$ with two independent random seeds $s_A = \left\{s_{A1}, s_{A2},\cdots, s_{Am} \right\}$, $s_B = \left\{s_{B1}, s_{B2},\cdots, s_{Bn}\right\}$. $m$ and $n$ are determined by the min-entropy and the length of raw data $l$. And $s_A$, $s_B$ consisted the row and column of $T$, respectively. 
\begin{align}
n &= l \notag \\
m &= l\times \frac{H_m}{H_0}-2log_2(\epsilon)
\end{align}
Notice that, $\epsilon$ is the secure parameter, and $H_0 = log_2(l)$, $H_m$ is the min-entropy of the raw data.

\textit{Scheme of postprocessing---} For a raw data sequence $d$ with length $l$, eligible quantum random sequence $d^{'}$ can be obtained as follows:
\begin{align}
&d \times T = d^{'} \notag \\
\underbrace{\left(d_{1}, \cdots, d_{l}\right)}_\text{l} &\times 
\underbrace{\left(\begin{array}{ccc} s_{A1} & \cdots & s_{Am} \\  \vdots & \ddots & \vdots \\ s_{Bn} & \cdots & s_{A1} 
\end{array}\right)}_{m} = \left(d^{'}_{1}, \cdots, d^{'}_{m}\right) 
\end{align}

Here, we noticed a systematic bias ascribed to the low peak-peak value of our QRNG. Briefly, the lower level in our experiment was supposed to be low enough so that no tunneling could occur and the after-pulse could be relieved. Unfortunately, restricted by the inner set of the driven module in InGaAs/InP trigger module, the difference between high level and low level is fixed at $\Delta U = 4V$, which means that even the low level$U_L = U_H - 4V$, and could result in tunneling current. Inevitably, the InGaAs/InP APD self-protecting program, which compels the InGaAs/InP APD out of avalanche effect, was activated automatically as the InGaAs/InP APD works at avalanche mode in a prolonged period. Thus, we could see a set of $0$s in our raw data periodically.

\section{\uppercase\expandafter{\romannumeral5}.Data analysis}

In our final experiment, we set the frequency of trigger pulse to $50MHz$, with $0ns$ deadtime. And the bias voltage was fixed to $49.40V$. As showed above, the min-entropy $H_m = 5.12$, for $8-bits$ sequences. Combined with a secure parameter $\epsilon = 2^{-100}$, data length $l = 3000$, the random bits generation rate was $8.3Mb/s$.

After a set of $5Gb$final data, we applied the $NIST-sts$ Test and $Diehard$ Test. See Figure\ref{fig5} and Figure\ref{fig6}. Aside from these two tests, the auto-correlation function is also drew from the data sheet, as shown by Figure\ref{fig7}. And more details of the test data is provided in the $Appendix$.

\begin{figure}
\includegraphics[width=3.5in]{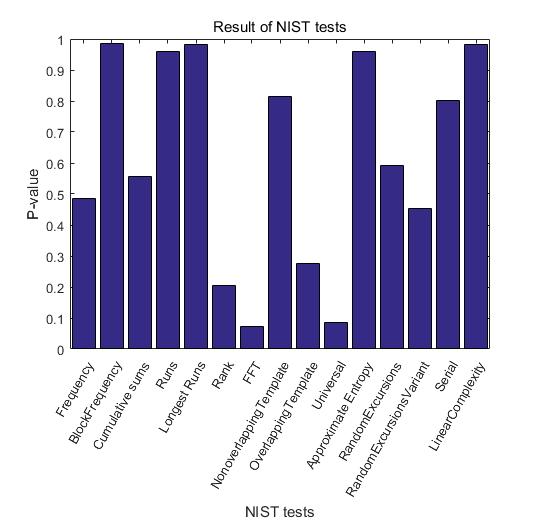}
\caption{Upper is the result of NIST test of our final random sequence while the lower is the eligible rate of sequences decomposed from the original final sequence. The voltage of high level $V_h = 49.40V$ and the data size of original final sequence is 5Gb.}
\label{fig8}
\end{figure}

\begin{figure}
\includegraphics[width=3.5in]{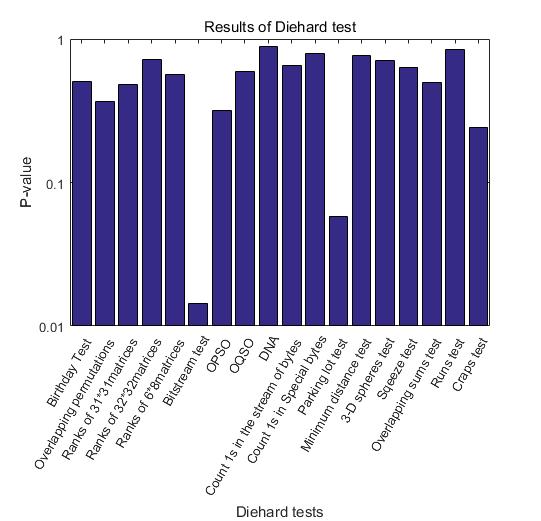}
\caption{Upper is the result of Diehard test of our final random sequence . The voltage of high level $V_h = 49.40V$ and the data size of original final sequence is 5Gb.}
\label{fig9}
\end{figure}

\begin{figure}
\includegraphics[width=3.5in]{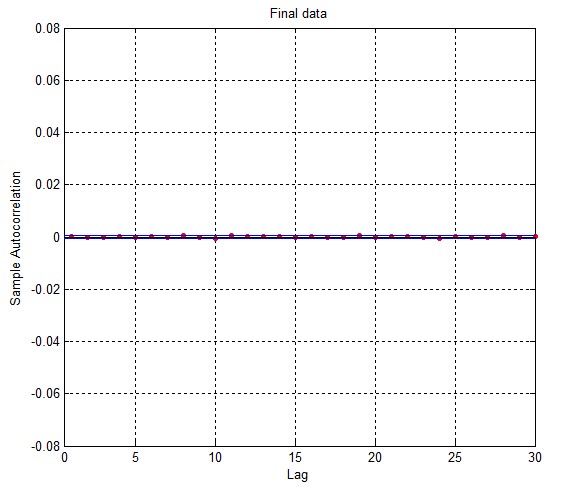}
\caption{Auto-correlation of the forgoing final data.}
\label{fig10}
\end{figure}

It is apparently that all our random number passed these two tests and the auto-correlation got a dramatic decline after postprocessing with Toeplitz-hashing extractor. Noteworthy, that the improvement on the data collection module and the optimization of the trigger module is approachable certifies realization of higher generation rate.
\section{\uppercase\expandafter{\romannumeral6}. Conclusion}

In this paper, we proposed a QRNG protocol based on the tunneling effect in InGaAs/InP avalanche diode. And thus no photonic sources is required in our experiments. Moreover, with the application of integrated module in InGaAs/InP single photon detector, we implemented a photonic-source-free QRNG, whose generation rate could reach $8.3Mb/s$. Moreover, this rate can be lifted up to $20Mb/s$ with the facilities we have. Our further study will focus on following questions:

\textit{1. } Designing a trigger source with higher frequency, shorter pulse width and larger peak-peak voltage value.

\textit{2. } Seeking a more stable and robust physical system as the tunneling source, in light of the disadvantages of our InGaAs/InP avalanche diode system.

\textit{3. } Combination of this tunneling protocol and other QRNGs, as mentioned in \cite{lunghi2015self}\cite{PhysRevX.6.011024}\cite{xu2016experimental}\cite{ma2015quantum}.

\section{ Acknowledgments}

We acknowledge Weixing Zhang and Hua Yuan for their assistance on the hardware designing. And we thank Pro.Xiongfeng Ma and Dr.Zhen Zhang for crucial discussions and Xinyu Liu, Nan Jiang for their guidance on the application of several sets of test software. Also, we thank the NSFC for its financial support.

\section{ Appendix: Detailed result of randomness tests}
The detailed data analysis by NIST test is obtained by the official program 'sts' version $2.1.2$, as shown in TABLE\ref{tab1} . And the detailed data analysis by Dihard test is shown as the following TABLE\ref{tab2}:

\begin{table}
\centering
\begin{tabular}{lccc}\hline Statistical Test &P-value & Proportion & Assessment\\\hline Frequency & 0.484838 & 0.993355 & Success\\ BlockFrequency & 0.984047 & 0.986711 & Success \\
CumulativeSums & 0.557001 &0.993355 & Success\\ Runs & 0.958728 & 0.996678 & Success\\ LongestRun &0.981358&0.983389 & Success\\ Rank & 0.204974 &0.986711 & Success\\ FFT & 0.071125 &0.986711 & Success\\NonOverlappingTemplate &0.814243 & 0.976744 & Success\\OverlappingTemplate & 0.274627 & 0.993355 & Success\\ Universal & 0.084015 & 0.983389  & Success\\ ApproximateEntropy & 0.959407 & 1.000000 & Success\\RandomExcursions & 0.592779 & 0.977011 & Success\\RandomExcursionsVariant & 0.452699 & 0.977011 & Success \\Serial & 0.800792 & 0.990033 & Success\\ LinearComplexity & 0.982786 & 0.986711 & Success \\\hline 
\end{tabular}
\label{tab1}
\caption{Result of NIST test for a $5Gb$ final data, The minimum pass rate for each statistical test with the exception of the random excursion (variant) test is approximately = 292 for a sample size = 301 binary sequences. The minimum pass rate for the random excursion (variant) test is approximately = 168 for a sample size = 174 binary sequences. As the confidence parameter $\alpha = 0.01$, our data passed the NIST test. }
\end{table}

\begin{table}
\centering
\begin{tabular}{lcc}\hline StatisticalTest & P-value & Assessment\\\hline BirthdayTest & 0.505898 & Success\\ OverlappingPermutation & 0.368835 & Success \\RanksOf$31\times 31$matrices & 0.481990 & Success\\ RanksOf$32 \times 32$matrices & 0.714278 & Success\\ RanksOf$6 \times 8$matrices &0.566601 & Success\\ BitstreamTest &0.01443 & Success\\ OPSO &0.317900 & Success\\OQSO &0.592000 & Success\\DNA  & 0.883100 & Success\\ Count$1s$inTheStreamOfBytes& 0.648895  & Success\\ Count $1s$InTheSpecialBytes & 0.790766 & Success\\ParkingLotTest & 0.058110 & Success\\MinimumDistanceTest & 0.762900 & Success \\$3-D$ SpheresTest &0.705579 & Success\\ SqueezeTest& 0.634944 & Success \\OverlappingSumsTest & 0.501220& Success\\ Runs & 0.846631& Success \\ Craps & 0.242872& Success\\\hline

\end{tabular}
\caption{Result of Diehard test for a $5Gb$ final data, All of these indexes lies in$(0,1)$, our data passed the Diehard test. }
\label{tab2} 
\end{table}

\vskip 12pt
{\fontsize{7.8pt}{9.4pt}\selectfont

\end{document}